\documentclass[number,sort&compress,12pt,3p,preprint,times]{elsarticle}
\usepackage[draft]{hyperref}
\usepackage{graphicx}
\usepackage{amsmath}
\usepackage{amssymb}
\usepackage{tabularx}
\usepackage{multirow}
\usepackage{ulem} %strikout
\usepackage{cleveref}%clever references. no need to handle Figure or Table in text
\usepackage[usenames,dvipsnames,svgnames,table]{xcolor}

\newcommand{\veritas}{{VERITAS}}
\newcommand{\hess}{{H.E.S.S.}}
\newcommand{\magic}{{MAGIC}}
\newcommand{\fact}{{FACT}}
\newcommand{\clue}{{CLUE}}
\newcommand{\cta}{{CTA}}
\newcommand{\hegra}{{HEGRA}}
\newcommand{\whipple}{{Whipple}}
\renewcommand{\deg}{$^\circ$}
\newcommand{\ie}{{i.e.}}
\newcommand{\GeV}{\mathrm{GeV}}

\renewcommand{\cref}{\Cref}

\newcommand{\captionfonts}{\normalsize}
\makeatletter  % Allow the use of @ in command names
\long\def\@makecaption#1#2{%
  \vskip\abovecaptionskip
  \sbox\@tempboxa{{\captionfonts #1: #2}}%
  \ifdim \wd\@tempboxa >\hsize
    {\captionfonts #1: #2\par}
  \else
    \hbox to\hsize{\hfil\box\@tempboxa\hfil}%
  \fi
  \vskip\belowcaptionskip}
\makeatother   % Cancel the effect of \makeatletter

\makeatletter
\long\def\MaketitleBox{%
  \resetTitleCounters
  \def\baselinestretch{1}%
  \begin{center}%
   \def\baselinestretch{1}%
    \Large\@title\par\vskip18pt
    \normalsize\elsauthors\par\vskip10pt
    \footnotesize\itshape\elsaddress\par\vskip36pt
    \clearpage
    \hrule\vskip12pt
    \ifvoid\absbox\else\unvbox\absbox\par\vskip10pt\fi
    \ifvoid\keybox\else\unvbox\keybox\par\vskip10pt\fi
    \hrule\vskip12pt
    \end{center}%
  }
\makeatother

\journal{Astroparticle Physics}

\begin{document}

\begin{frontmatter}
\title{Gamma-ray Observations Under Bright Moonlight with VERITAS}
\author[1]{S.~Archambault}
\author[2]{A.~Archer}
\author[3]{W.~Benbow}
\author[4,21]{R.~Bird\corref{cor1}}
\ead{ralphbird@astro.ucla.edu}
\author[1]{E.~Bourbeau}
\author[5,AB]{A.~Bouvier}
\author[4]{M.~Buchovecky}
\author[2]{V.~Bugaev}
\author[6]{J.~V~Cardenzana}
\author[3]{M.~Cerruti}
\author[7]{L.~Ciupik}
\author[8]{M.~P.~Connolly}
\author[9,10]{W.~Cui}
\author[3]{M.~K.~Daniel}
\author[2]{M.~Errando}
\author[11]{A.~Falcone}
\author[1]{Q.~Feng}
\author[9]{J.~P.~Finley}
\author[12]{H.~Fleischhack}
\author[13]{L.~Fortson}
\author[14]{A.~Furniss}
\author[8]{G.~H.~Gillanders}
\author[1]{S.~Griffin\corref{cor1}}
\ead{griffins@physics.mcgill.ca}
\author[1]{D.~Hanna}
\author[5]{O.~Hervet}
\author[15]{J.~Holder}
\author[3]{G.~Hughes}
\author[16]{T.~B.~Humensky}
\author[12]{M.~H{\"u}tten}
\author[5]{C.~A.~Johnson}
\author[17]{P.~Kaaret}
\author[18]{P.~Kar}
\author[19]{M.~Kertzman}
\author[18]{D.~Kieda}
\author[12]{M.~Krause}
\author[8]{M.~J.~Lang}
\author[1]{T.~T.~Y.~Lin}
\author[12]{G.~Maier}
\author[9]{S.~McArthur}
\author[8]{P.~Moriarty}
\author[20]{R.~Mukherjee}
\author[16]{D.~Nieto}
\author[21]{S.~O'Brien}
\author[4]{R.~A.~Ong}
\author[22]{A.~N.~Otte}
\author[23]{N.~Park}
\author[24,12]{M.~Pohl}
\author[4]{A.~Popkow}
\author[21]{E.~Pueschel}
\author[21]{J.~Quinn}
\author[1]{K.~Ragan}
\author[25]{P.~T.~Reynolds}
\author[22]{G.~T.~Richards}
\author[3]{E.~Roache}
\author[13]{C.~Rulten}
\author[12]{I.~Sadeh}
\author[9]{G.~H.~Sembroski}
\author[13]{K.~Shahinyan}
\author[23]{D.~Staszak}
\author[24,12]{I.~Telezhinsky}
\author[1]{S.~Trepanier}
\author[23]{S.~P.~Wakely}
\author[6]{A.~Weinstein}
\author[17]{P.~Wilcox}
\author[5]{D.~A.~Williams}
\author[1]{B.~Zitzer}
\cortext[cor1]{Corresponding authors}
\address[1]{Physics Department, McGill University, Montreal, QC H3A 2T8, Canada}
\address[2]{Department of Physics, Washington University, St. Louis, MO 63130, USA}
\address[3]{Fred Lawrence Whipple Observatory, Harvard-Smithsonian Center for Astrophysics, Amado, AZ 85645, USA}
\address[4]{Department of Physics and Astronomy, University of California, Los Angeles, CA 90095, USA}
\address[21]{School of Physics, University College Dublin, Belfield, Dublin 4, Ireland}
\address[5]{Santa Cruz Institute for Particle Physics and Department of Physics, University of California, Santa Cruz, CA 95064, USA}
\address[AB]{InVisage Technologies, Inc., Menlo Park, CA 94025, USA}
\address[6]{Department of Physics and Astronomy, Iowa State University, Ames, IA 50011, USA}
\address[7]{Astronomy Department, Adler Planetarium and Astronomy Museum, Chicago, IL 60605, USA}
\address[8]{School of Physics, National University of Ireland Galway, University Road, Galway, Ireland}
\address[9]{Department of Physics and Astronomy, Purdue University, West Lafayette, IN 47907, USA}
\address[10]{Department of Physics and Center for Astrophysics, Tsinghua University, Beijing 100084, China.}
\address[11]{Department of Astronomy and Astrophysics, 525 Davey Lab, Pennsylvania State University, University Park, PA 16802, USA}
\address[12]{DESY, Platanenallee 6, 15738 Zeuthen, Germany}
\address[13]{School of Physics and Astronomy, University of Minnesota, Minneapolis, MN 55455, USA}
\address[14]{Department of Physics, California State University - East Bay, Hayward, CA 94542, USA}
\address[15]{Department of Physics and Astronomy and the Bartol Research Institute, University of Delaware, Newark, DE 19716, USA}
\address[16]{Physics Department, Columbia University, New York, NY 10027, USA}
\address[17]{Department of Physics and Astronomy, University of Iowa, Van Allen Hall, Iowa City, IA 52242, USA}
\address[18]{Department of Physics and Astronomy, University of Utah, Salt Lake City, UT 84112, USA}
\address[19]{Department of Physics and Astronomy, DePauw University, Greencastle, IN 46135-0037, USA}
\address[20]{Department of Physics and Astronomy, Barnard College, Columbia University, NY 10027, USA}
\address[22]{School of Physics and Center for Relativistic Astrophysics, Georgia Institute of Technology, 837 State Street NW, Atlanta, GA 30332-0430}
\address[23]{Enrico Fermi Institute, University of Chicago, Chicago, IL 60637, USA}
\address[24]{Institute of Physics and Astronomy, University of Potsdam, 14476 Potsdam-Golm, Germany}
\address[25]{Department of Physical Sciences, Cork Institute of Technology, Bishopstown, Cork, Ireland}

\begin{abstract}
Imaging atmospheric Cherenkov telescopes (IACTs) are equipped with sensitive photomultiplier tube (PMT) cameras. 
Exposure to high levels of background illumination degrades the efficiency of and potentially destroys these photo-detectors over time, so IACTs cannot be operated in the same configuration in the presence of bright moonlight as under dark skies.
Since September 2012, observations have been carried out with the \veritas\ IACTs under bright moonlight (defined as about three times the night-sky-background (NSB) of a dark extragalactic field, typically occurring when Moon illumination $> 35\%$) in two observing modes, firstly by reducing the voltage applied to the PMTs and, secondly, with the addition of ultra-violet (UV) bandpass filters to the cameras.  
This has allowed observations at up to about 30 times previous NSB levels (around $80\%$ Moon illumination), resulting in $30\%$ more observing time between the two modes over the course of a year.  
These additional observations have already allowed for the detection of a flare from the 1ES~1727+502 and for an observing program targeting a measurement of the cosmic-ray positron fraction.
We provide details of these new observing modes and their performance relative to the standard \veritas\ observations.
\end{abstract}
\end{frontmatter}

%Section heading
\section{Introduction}
Imaging atmospheric Cherenkov telescopes (IACTs) are the predominant instruments for very-high-energy (VHE, $100~\GeV < E < 30~\mathrm{TeV}$) gamma-ray astrophysics, in particular for high-resolution, high-sensitivity observations.  
However, most IACTs use photomultiplier tubes (PMTs) in their cameras, limiting the background light levels under which they can operate. 
If sufficiently high this background light can degrade or damage the PMTs.  
This presents a limitation on the amount of observing time available for source discovery or deep exposures.  In particular this impacts studies of varying or transient sources where a regular cadence of observations is desirable.

Gamma rays or cosmic rays incident on the Earth's atmosphere generate a relativistic particle cascade, known as an ``extensive air shower''. 
IACTs detect these showers through the Cherenkov radiation induced by the particles as they propagate through the atmosphere. 
The mostly blue and UV Cherenkov light arrives at the ground tightly bunched in time (within a few nanoseconds) and spread over a roughly circular area of radius $\sim$~120~m. The amount of Cherenkov light produced is proportional to the energy of the shower primary; Cherenkov photon intensities range from a few $\mathrm{photons/m^{2}}$ at~100~GeV to over 10,000 $\mathrm{photons/m^{2}}$ at 100~TeV \citep{2003vheg.book.....W}.
In contrast, the night-sky-background (NSB) at the \veritas\ site for wavelengths 290~nm $<$ $\lambda$ $<$ 650~nm and a dark, extragalactic field (elevations $>$~60\deg, outside the galactic plane and excluding bright stars) is ($3.2 \pm 0.8) \times 10^{12}~\mathrm{ph/sr/s/m}^2$ \citep{2008PhDT........65V}. Under a full moon, the NSB increases by a factor of 100 in the $U$-band (which is most relevant to Cherenkov telescopes) \citep{Elias1994}.
This NSB light provides a background above which the Cherenkov light must be detected.
The energy threshold of an IACT is determined by the ability to image faint showers above the NSB and imaging these showers can dramatically increase the sensitivity of an IACT to many sources due to the power law nature of both the the source and cosmic ray background spectra.

By using the atmosphere as part of the detector in this way, the telescopes are able to image showers that fall over a large area, resulting in an effective collection area of $\sim10^5~\mathrm{m^2}$, orders of magnitude larger than the actual mirror area. 
An overview of the atmospheric Cherenkov technique is given in \cite{2003vheg.book.....W}.

Photo-multiplier tubes (PMTs) are the default photosensors for IACT cameras due to their high gain, low noise and fast response.
They are used by all three of the major IACT arrays currently in operation: \hess\ \citep{2006A&A...457..899A}, \magic\ \citep{2008ApJ...674.1037A}, and \veritas\ \citep{2002APh....17..221W}.
The major disadvantage of PMTs is that exposure to bright light, even without the application of a high voltage, can result in damage.
PMTs deteriorate over time due to the aging of their dynodes by the impact of electrons during the multiplication process.  
Aging occurs at a rate correlated with the integrated anode current, which is increased when NSB levels are elevated.
For this reason, the \veritas\ Collaboration imposes upper limits on the mean anode current (which depends on the NSB level) at 15~$\mu$A\footnote{A higher limit exists for individual pixels to accommodate stars in the field-of-view. 
Unless the telescope is pointed directly at a star, the star will only remain in any part of the camera for a relatively short period of time due to the fact that the star field rotates as the telescope tracks a given point on the sky.}. 
This level allows for multiple years of stable operation; the slow degradation of the PMT dynodes due to this current is compensated for by adjusting the PMT gains on a regular basis \cite{2011arXiv1110.4702N}.
For comparison, the mean anode current for a dark, extragalactic field is 5~$\mu$A.
Operating at higher currents results in faster degradation, requiring more frequent gain changes and reducing the overall lifetime of the PMTs.  
In addition, higher levels of NSB also reduce the signal-to-noise ratio of the shower Cherenkov light over the NSB. 
This reduces the sensitivity of the IACT, as well as raising the energy threshold. 
For \veritas, this maximum anode current limits the observing time per year to about 1800 hours. 
Adverse weather conditions preventing observations, the \veritas\ summer shutdown due to Arizona monsoons, and other losses reduce the number of observing hours to approximately 1000~hours per year.

The motivation for this work is to reduce the impact of NSB on the operation of \veritas, thus increasing the duty cycle of the experiment.  
Though the relationship between anode current and Moon illumination is complex, depending on relative location in the sky and atmospheric conditions, the changeover to the new observing modes presented in this paper typically occurs when the Moon illumination exceeds about 35\%.
Any observations that are conducted above the NSB limit are described as ``bright'' moonlight observations as without altering the functioning of the array they would not be possible.
In this paper we describe how we have conducted observations above this threshold at the cost of sensitivity or energy threshold, to allow deeper exposures (in particular at high energies where sensitivity is minimally affected by the NSB) and increased and more regular monitoring of variable sources.
 
\section{History of Moonlight Observations}
Several techniques have been developed to reduce the impact of sky brightness on IACTs.
These can loosely be broken into three categories, described below.

\subsection{Lowering the PMT gain}
In order to reduce the damage to the dynodes it is possible to lower the gain of a PMT (\ie\ the amplification factor of a signal produced at the photocathode) by lowering the high voltage applied to the PMT; this is henceforth referred to as ``reduced high voltage'' (RHV) operation.  
This technique was pioneered by the \hegra\ Collaboration
\citep{1999APh....12...65K} where they used a variety of voltage levels to observe a number of sources, in particular bright, variable blazars such as 1ES~1959+650 \citep{2003ICRC....5.2615T} and Mrk~421 \citep{2001ICRC....7.2687C}.  
RHV observations under moonlight suffer from lower signal-to-noise due to the increased NSB levels; as a result, sensitivity to faint or distant showers is reduced and the effective energy threshold is increased. 
This reduces the effective collecting area of the array.

The \magic\ Collaboration uses PMTs that operate at lower gains than those employed by the \veritas\ Collaboration ($3\times 10^4 $ vs. $2\times10^5$); this enables them to operate under  moderate moonlight and twilight conditions without needing to lower the PMT gains to accommodate the higher light levels \citep{2007astro.ph..2475M,2009arXiv0907.0973B}.

\subsection{Reduced optical sensitivity}
An alternative approach has been to develop a technique to reduce the optical sensitivity of the camera, in particular to shape the spectral response of the camera to favor the UV part of the electromagnetic spectrum  where most Cherenkov light is emitted.  
By using a camera that is sensitive to the UV, but not the optical, it is possible to reduce the noise from the NSB significantly more than the signal from the Cherenkov light, improving the signal-to-noise ratio while also protecting the camera against the increased NSB.

The \whipple\ Collaboration tested solar-blind PMTs and a liquid UV filter to conduct operations under moderate moonlight \citep{1986BAAS...18..700W,1995ICRC....2..544C,1996NIMPA.368..503U}.
The liquid filter was contained between two quartz plates and increased the sensitivity of the experiment to air showers by a factor of three over the NSB.

Recently, the \magic\ collaboration has begun using UV filters to extend their observations through the full lunar cycle \citep{2015arXiv150902048G}. 

\subsection{Alternative camera types}
An alternative approach was taken by the \clue\ Collaboration which built an array of nine telescopes on the island of La Palma, Spain. 
To achieve their desired UV sensitivity they used a multiwire proportional chamber filled with TMAE (tetrakis(dimethylamino)ethylene) in place of PMTs.  
This substance has a particularly good photoconversion in the $180-240~\mathrm{nm}$ range.  
The \clue\ array was able to detect strong gamma-ray sources (the Crab Nebula and Mrk~421) as well as the deficit in the cosmic ray flux in the direction of the Moon \citep{Bastieri2000,Dokoutchaeva2001}.

Recent advances in silicon photomultiplier (SiPM) development have led to their use in cameras of IACTs. 
They are currently used on the First G-APD Cherenkov Telescope (\fact) \citep{2013JInst...8P6008A} and are planned for use in some telescope types in the upcoming Cherenkov Telescope Array (\cta) \citep[e.g.][]{2015arXiv150902345O, 2013arXiv1307.2807D,2015arXiv150806453S,2015arXiv150901979A,2015arXiv150807120R}. 
One of the main advantages of SiPMs is that they are not damaged  by operation under very high background light conditions.  
This enables safe operation even under a full Moon \citep{2013arXiv1307.6116K}.  
However, SiPMs do not overcome the increase in background light and this, in turn, reduces the sensitivity to small showers, which increases the energy threshold and decreases the effective area. 
Their significant sensitivity to red light also makes them more susceptible to these effects than PMTs.

\section{The Very Energetic Radiation Imaging Telescope Array System (VERITAS)}
\label{sec:VERITAS}
\veritas\ is an array of four imaging atmospheric Cherenkov telescopes, located at the Fred Lawrence Whipple Observatory in southern Arizona.  
Each telescope is of Davies-Cotton design, with a 12-m aperture reflector comprising 350 hexagonal mirror facets.  
The focal length of each telescope is 12~m and each is equipped with a camera consisting of 499 close-packed photomultiplier tube ``pixels'' at the focus.  
The angular spacing between PMTs is 0.15\deg\ which yields a total field of view of 3.5\deg\ in diameter.  
A Winston-type light cone is mounted in front of each PMT to reduce the dead space between pixels, to reject stray light not arriving from the telescope reflector and to increase the light collection efficiency at the photocathodes from $55\%$ to $75\%$ \citep{2007arXiv0709.4517V}.
LED flashers are used for pixel calibration \citep{2010NIMPA.612..278H} and short (2~-~5 minute) dedicated calibration runs are conducted each night for each of the observing modes employed.
\veritas\ began full array operations in 2007 and in summer 2009 the first \veritas\ telescope was relocated to increase the sensitivity of the array \citep{2009arXiv0912.3841P}.

\veritas\ uses a three-level trigger system for data acquisition \citep{2008ICRC....3.1539W}.  
The first level (L1) uses a programmable constant-fraction discriminator (CFD) requiring a PMT pulse height above a given threshold. 
This is set using a bias curve (\Cref{fig:BiasCurves}), at lower thresholds the triggers are dominated by NSB photons whereas at higher thresholds the trigger rate is dominated by light from cosmic ray showers.  
The CFD level is set to a value close to the inflection point of the bias curve to maximize the amount of useful data recorded; for normal observations the threshold is 45~mV, (around 4-5 photoelectrons). 
The turnover at very high rates ($>$1~MHz)  is due to saturation in the \veritas\ trigger system.
The second level (L2, also known as the pattern trigger) requires a coincidence between three adjacent PMTs within a given coincidence window (currently 5~ns).  
The third level (L3, or array trigger) requires an L2 trigger from at least 2 telescopes within a 50~ns coincidence window.  
PMT signals are continuously digitized by 500 mega-samples per second flash analogue-to-digital converters (FADCs) to a memory buffer.  
If the L3 trigger criteria are met, 32~ns of data is read out from this memory buffer and stored to disk.  
In 2011, \veritas\ upgraded the L2 trigger to a field-programmable gate array (FPGA) based system.  
This new system allows for tighter coincidence windows between adjacent PMTs which reduces accidental L2 (and thus accidental L3) triggers \citep{2013arXiv1307.8360Z}. 

In summer 2012 the cameras in each telescope were upgraded with new, high quantum efficiency photomultiplier tubes (from Photonis XP2970 with a peak quantum efficiency of about 23\% at 400~nm to Hamamatsu R10560-100-20MOD, with a peak quantum efficiency of about 35\% at 350~nm) \citep{2011arXiv1110.4702N}.
This has resulted in a decrease of the analysis energy threshold to $\sim$~85~GeV from $\sim$~130~GeV for studies targeting low-energy processes (e.g. pulsar analyses optimized for sources with very soft spectra).

\section{VERITAS Observations Under Moonlight}
\subsection{Raised CFD Trigger Threshold Observations}
\begin{figure}[htb!]
\centering
\includegraphics[width=\linewidth]{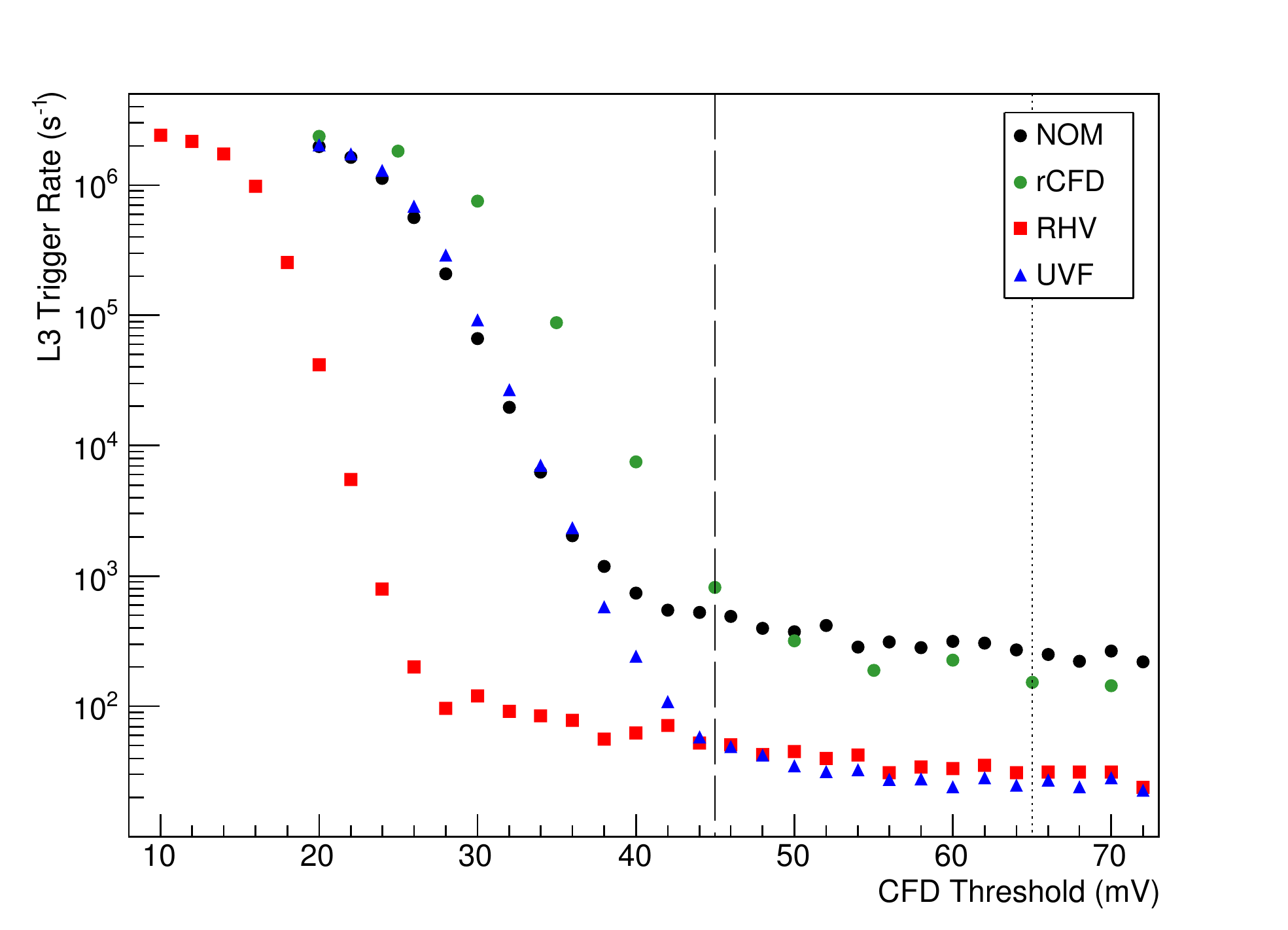}
\caption{
Array trigger rate versus CFD threshold for four different observing modes under typical conditions for that mode. 
``rCFD'' corresponds to data taken under low levels of moonlight when observations are conducted with raised CFD thresholds at 65~mV (dotted line) rather than the normal 45~mV (dashed line).
It can clearly be seen that for low thresholds, the trigger rate is substantially higher even under low amounts of moonlight, in this case raising the threshold from 45 to 65~mV results in a reduction in the L3 rate by a factor of $\sim 5$ for the rCFD data. 
Data for the RHV and UVF observing modes (described later in the main text) are also shown.
\label{fig:BiasCurves}}
\end{figure}

Under low levels of moonlight (typically with the lunar disk less than $35\%$ illuminated or when the moon is close to the horizon or is obscured by nearby mountains) the PMT currents do not rise to a level which would be damaging.
However, the accidental trigger rate does increase (see \cref{fig:BiasCurves}), increasing the array deadtime and reducing the overall sensitivity of the experiment.  
To counteract this, the CFD thresholds are raised from 45 to $65~\mathrm{mV}$, reducing the accidental trigger rate at the cost of sensitivity to the dimmest showers.  
The CFD threshold change is implemented when the average currents reach 10~$\mu$A, about twice the level of a dark extra-galactic field and above all but the brightest galactic fields.
This method has been used for many years by \veritas\ and was first used in the detection of W Comae \citep{Acciari2009a}.
Note that the location of the inflection point changes as a function of many parameters (moon illumination, elevation, and angle from telescope pointing, etc.) so in the case of the data taken under low amounts of moonlight (``rCFD'') shown in \cref{fig:BiasCurves} the inflection point is below 65~mV.
65~mV has been chosen conservatively to allow observations to take place up to currents of 15~$\mu$A while maintaining a sufficiently low L3 rate that the array deadtime remains low.

\subsection{Reduced High Voltage Observations}
By reducing the high voltage applied to a PMT its gain is lowered, reducing the current through the dynodes and reducing the rate of damage.
However, though the reduction in the gain protects the dynodes, it does not provide any preferential reduction in the NSB over the Cherenkov signal; thus it becomes harder to discriminate weak signals from the increased background. 
This reduces the sensitivity to faint showers, increasing the energy threshold.  
It is therefore necessary to choose a level of reduction that provides a suitable compromise between the requirements: reducing the gain sufficiently to protect the camera, while also operating the PMTs within specifications and maintaining maximum sensitivity to faint showers. 
This prevents further reductions in the high voltage.

The PMTs used by \veritas\ have a gain dependence on high voltage that can be described by a power law of index approximately six. 
When operating in RHV mode, the PMTs are operated at a gain of about 30\% of normal (which corresponds to 81\% of the nominal voltage).
To correct for PMT-to-PMT variation in the precise values of the power-law index, a flat-fielding procedure is performed at the lower voltage to improve the uniformity of the camera's response (as described in \Cref{sec:VERITAS}).  
As with normal running, the CFD thresholds are adjusted according to the prevailing NSB conditions, typically to 25~mV (see \Cref{fig:BiasCurves}).
Due to the low trigger rates in RHV mode it is possible to operate below the inflection point without incurring a significant increase in the array deadtime and use offline analysis to remove NSB triggers.  
This has the advantage of lowering the trigger energy threshold for RHV observations.

Operating in the RHV mode allows \veritas\ to conduct observations when the average currents rise above 15~$\mu$A (which corresponds to an NSB level about three times a dark extragalactic field and typically occurs when the Moon reaches 35\% illumination). 
Observations continue in RHV mode until the currents again reach the upper limit of 15~$\mu$A (about 10 times the NSB level of a dark, extragalactic field, around 65\% Moon illumination).
To continue observing above this level would require a further reduction in the high voltage.
This is not possible while continuing to operate the PMTs within specifications, thus \veritas\ uses an alternate observing mode, described next.

\subsection{UV Filter Observations}
\begin{figure}[hbt!]
\centering
\includegraphics[width=\linewidth]{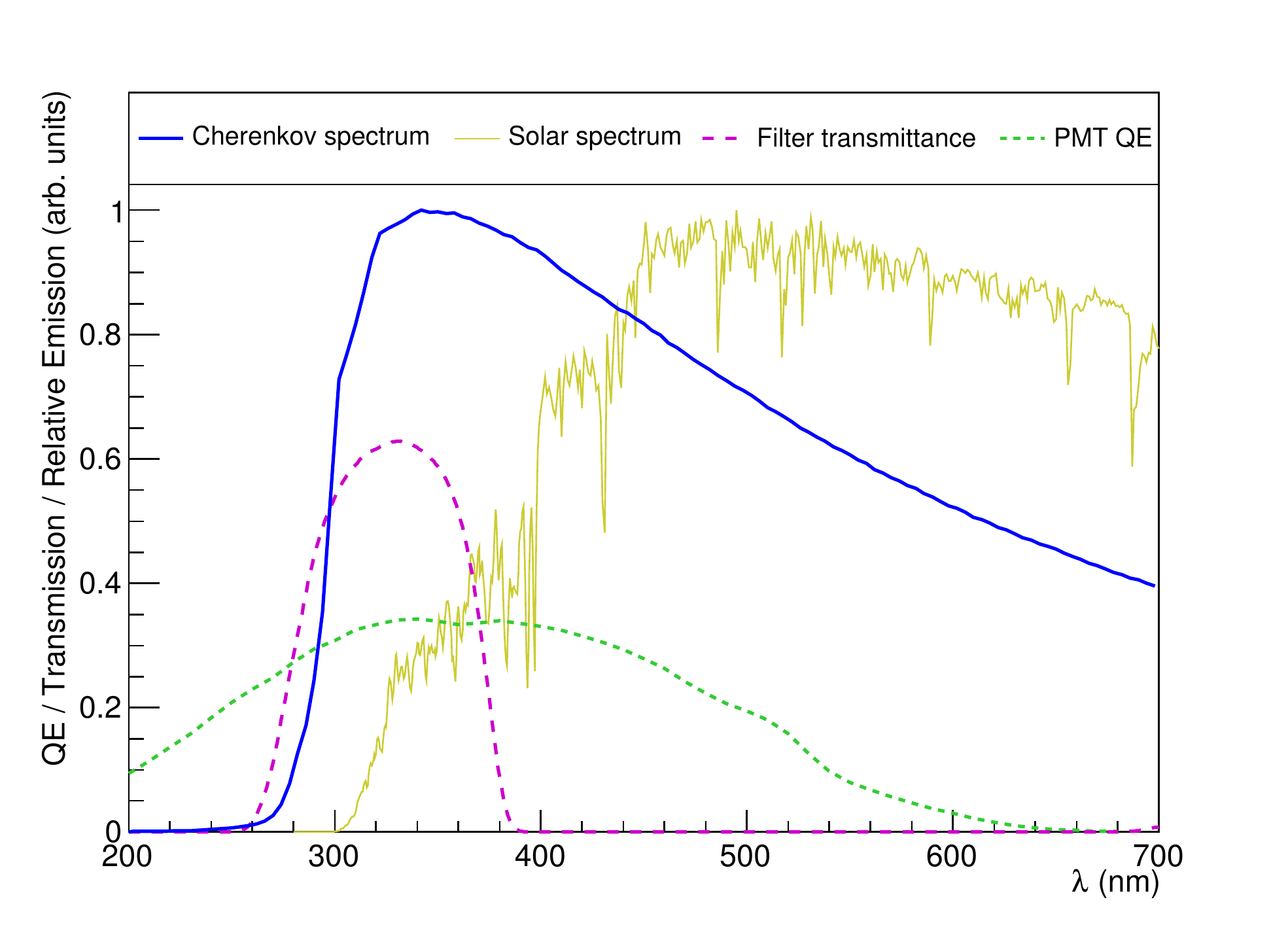}
\caption{\veritas\ PMT (Hamamatsu R10560-100-20MOD) quantum efficiency  and SCHOTT UG-11 total filter transmittance (for a 3~mm thick filter including reflection and correction for non-normal incidence).  Also plotted are the Cherenkov spectrum of a 500~GeV gamma-ray shower and the solar spectrum.  Both spectra have been corrected for the effects of atmospheric absorption but not for the effects of mirror reflectivity.}
\label{fig:UVFtransmission}
\end{figure}

When the sky is too bright to allow RHV observations, \veritas\ switches to a UV filter (UVF) observing mode which uses UV bandpass filters to reject moonlight photons. 
\veritas\ uses 3-mm-thick SCHOTT UG-11 \cite{SCHOTT} glass filters.  
The transmission spectrum of the filters is given in \Cref{fig:UVFtransmission} with the ASTM G173-03 reference solar spectrum \cite{ASTM} and the spectrum of Cherenkov photons from a 500~GeV gamma-ray shower as seen at the altitude of the \veritas\ telescopes after accounting for atmospheric absorption. 
The lunar spectrum at the ground is similar to the solar spectrum at the ground, with variations due to the lunar albedo; see \cite{2004Msngr.118...11P} for a comparison.
When the filters are used, the integrated response of a \veritas\ PMT to the Cherenkov spectrum given in \Cref{fig:UVFtransmission} is reduced to approximately $15\%$  of the nominal response whereas the response to the solar spectrum is reduced to approximately $3\%$, giving a factor of approximately five improvement in the signal-to-NSB ratio as well as reducing the number of photons incident upon the PMTs.
The filters allow operation up to about 30 times the NSB of a dark, extragalactic field (about 80\% Moon illumination).

\begin{figure}[htb]
\centering
\includegraphics[width=\linewidth]{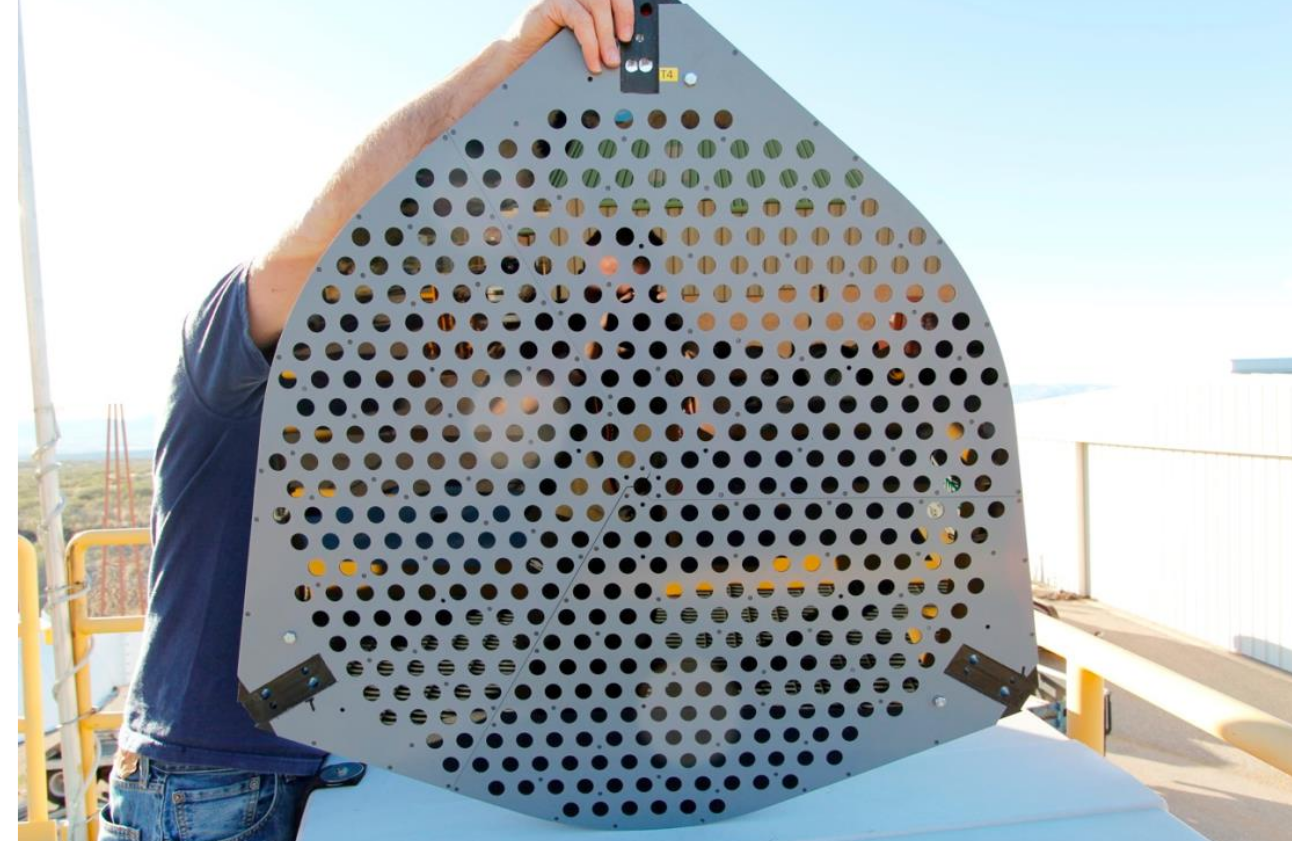}
\caption{A UV filter plate with one filter for each PMT mounted in a central plastic layer which is then sandwiched between two thin (gray) plastic sheets with holes smaller than the filters but larger than the exit apertures of the light cones.}
\label{fig:filters}
\end{figure}

Rather than one large piece of glass, which is fragile and expensive, a ``filter plate'' was manufactured using 499 individual filters (one for each PMT), as shown in \Cref{fig:filters}.  
The filter plate is made of three layers of plastic; the central layer has holes with the same diameter as the filters into which the filters are placed. 
This is then sealed between two thin sheets with holes smaller than the filters but larger than the exit apertures of the Winston cones (so there is no obstruction of the light exiting the cones).  
The plate is then mounted between the \veritas\ camera PMTs and light cones.  
The entire assembly is less than $5~\mathrm{mm}$ thick so it is a small perturbation to the optics of the camera assembly.

The filters have a nominal peak transmission of 72\% for normal-incidence photons \cite{SCHOTTUG11}, but this is reduced for photons entering at an angle, due to their increased path length in the filter and increased reflection from the filter surfaces.  
This reduction has been estimated using a ray-tracing program and has been determined to be approximately 15\%, bringing the peak transmission down to 62\% (\Cref{fig:UVFtransmission}).

Installation or removal of the filters by a pair of observers takes approximately 15 minutes per telescope, taking one hour for a typical observing crew of three.  
It should be noted that installation and removal are never done during normal observing (NOM) time.  
For example, one can install the filters during daylight hours and remove them before the Moon has set. 
After installation, as for normal observations, a flasher run is taken for gain correction, but no additional changes to the array settings are required (including the CFD thresholds, which are maintained at 45~mV, \Cref{fig:BiasCurves}).

\subsection{Bright Moonlight Observing  Program}

\begin{figure}[htb]
\centering
\includegraphics[width=\linewidth]{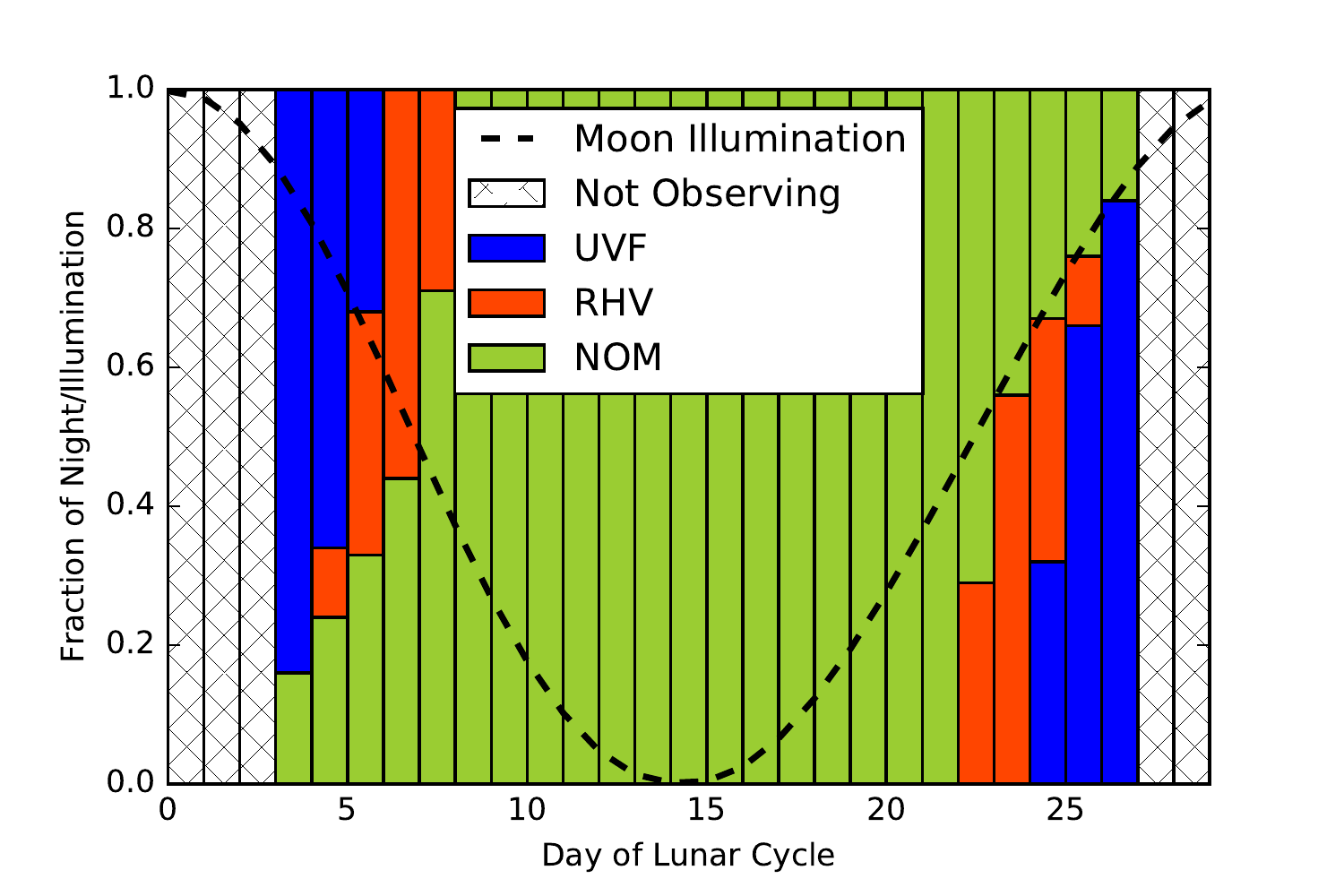}
\caption{Breakdown of a typical observing month for \veritas.  
This strategy is a guideline more than a rule; observing modes are chosen based on the PMT currents at the time of observing, rather than at fixed intervals.}
\label{fig:observingStrategy}
\end{figure}

Prior to the inception of the RHV and UVF observing modes, \veritas\ did not observe when the the average camera current rose above 15~$\mu$A.
The use of bright moonlight time has increased the number of hours \veritas\ can observe by roughly 30\%. 
The breakdown of a typical dark run (\ie\ a lunar cycle) is shown in \Cref{fig:observingStrategy}.  
As well as increased exposure for a large number of observing programs, the ability to observe under moonlight has enabled continued observations of transients through the so-called ``bright period'', increasing the chances of catching flaring objects and reducing the gaps in light curves.  
Note that the duration of the bright period and the breakdown between the different modes is dictated by the specific time at which the full moon occurs.  
Hence, the different modes can have slightly longer or shorter durations in different months.

When conducting observations with the Moon above the horizon, there is significant variation in the NSB level with angular distance from the Moon.  
The anode currents are lowest when the array is pointing about 90\deg\ from the Moon and the illumination levels quoted in this paper are based upon observations pointing in this direction.
A lack of suitable targets close to this location and/or the presence of clouds which reflect the moonlight and raise the anode currents reduces the Moon illumination at which the changeover between modes occurs.
It is noted that clouds have a significant impact on the anode currents, for this reason no data are taken in either RHV or UVF modes if there are clouds or any other adverse weather conditions which impact upon the quality of the data.

Due to the energy threshold and sensitivity differences between the observing modes, \veritas\ preferentially observes in the standard observing mode (NOM) rather than RHV mode, and RHV mode rather than UVF mode, wherever possible. 
This is done through target selection by the observers to select high elevation targets that lie as close as possible to 90\deg\ away from the Moon.
This maximizes the sensitivity of the instrument to gamma rays during moonlight observations and ensures the lowest-possible energy threshold.

\section{Analysis of Moonlight Data}
\label{Sec:Analysis}
Moonlight-data are analyzed using the standard \veritas\ analysis packages \citep{Daniel2007} but with the use of dedicated instrument response functions generated from simulations that incorporate the changes to the array.  
For the RHV simulations this is achieved by adjusting the gain in the detector model to the reduced value ($\sim30\%$ of nominal, as previously stated). 
UVF-mode simulations are produced using a detector configuration where the wavelength-dependent quantum efficiency of the PMTs is multiplied with the the wavelength-dependent transmission of the filters.

The detector response for each observing mode will be different for a given air shower, so analysis cuts must take this difference into consideration.  
In this work, the minimum image \textit{Size} (integrated signal in the image) requirement has been reduced by a factor of 3.5 for the analysis of RHV and UVF observations to account for the reduction in the \textit{Size} of the signal from the PMT for a given number of photons.
This reduction reflects the gain reduction for RHV observations ($1/3.5\approx 0.3$ corresponds to the $30\%$ gain reduction) with the aim of matching the analysis energy threshold between the NOM and RHV observing modes.
Though it would be possible to adjust the \textit{Size} cuts to match the analysis energy thresholds for all three observation types, this would increase the analysis energy threshold in comparison to that used in a typical \veritas\ analysis (due to the significantly higher trigger energy threshold of UVF observations).
For analyses requiring a lower energy threshold (e.g. soft-spectrum sources), it is possible to reduce the \textit{Size} cut for NOM observations to attain a lower energy threshold. 
However, RHV and UVF analyses already use a small \textit{Size} cut and thus the analysis energy thresholds in this work are close to the lower limit of that observing mode.

Other than the \textit{Size} cut, standard \veritas\ image-quality and gamma-hadron selection cuts were applied. 
These cuts were optimized on observations of the Crab Nebula with the signal strength scaled to $5\%$ of its measured value to simulate a weaker source using an independent, NOM data set. The aim of this optimization was maximizing the sensitivity rather than minimizing the analysis energy threshold. For this work, these standard cuts have been used to compare the performance of the different observation modes.  

It is possible to improve the sensitivity to hard-spectrum, weak sources by increasing the minimum \textit{Size} cut.
For example, for observations conducted in RHV mode, increasing the \textit{Size} cut by a factor of 3.5 to match that used in the NOM analysis reduces the time to detect a source with the strength of 1\% of the Crab Nebula from 1820 to 1200 minutes. 
Note that this improvement comes at the cost of a raised analysis energy threshold.  

\section{Sensitivity}
The observations of the Crab Nebula used in this work have been performed in each of the three observing modes (NOM, RHV and UVF), consisting of 320, 470, and 535 minutes of exposure, respectively (the NOM data set is a small subset of the \veritas\ data set chosen to have overall sensitivity and observing conditions comparable to the RHV and UVF data sets). 
All data were taken during good weather and with the source at high elevations (zenith angles $\lesssim 25^\circ$).  
The resulting sensitivity of each observing mode to sources with a Crab Nebula-like spectral shape and fluxes of various fractions of the Crab Nebula flux is given in
\cref{tab:SensitivityNumbers}, along with the energy threshold corresponding to each analysis (defined as the low energy edge of the lowest energy bin in the spectral reconstruction with an energy reconstruction bias $(E_{rec} - E_{MC}) / E_{MC}$ of less than 10\% for the Crab Nebula).  
In addition, \cref{tab:SensitivityNumbers} includes the sensitivity of each observing mode with a cut applied to remove showers with an energy below 1~TeV.
It can be seen that the RHV sensitivity and analysis energy threshold are very similar to the NOM sensitivity and analysis energy threshold with these cuts, while the UVF sensitivity is about a factor of two lower than the NOM and RHV sensitivity, this is in part due to the higher analysis energy threshold as can be seen in the smaller sensitivity difference above 1~TeV.

\begin{table*}
\centering
\begin{tabular}{cccccccc}
\hline\hline
     & \multicolumn{3}{c}{Time to Detection (min)} & Energy    & Raw Crab  & Corrected Crab  & Raw Crab Sensitivity \\
Mode & \multicolumn{3}{c}{Crab Flux Fraction}      & Threshold & Sensitivity & Sensitivity & (E \textgreater 1 TeV)  \\
     & 1             & 0.05         & 0.01         & (GeV)     & ($\sigma / \sqrt{\mathrm{hr}}$) & ($\sigma / \sqrt{\mathrm{hr}}$) & ($\sigma / \sqrt{\mathrm{hr}}$) \\ \hline
NOM  & 0.78          & 84.2         & 1760   & 158       & 43.9   & 47.9   & 17.2  \\
RHV  & 0.78          & 86.4         & 1820    & 158       & 43.8  & 45.4    & 16.3 \\
UVF  & 2.11          & 273          & 5980    & 251       & 26.7  &  26.9   & 14.2 \\ \hline\hline
\end{tabular}
\caption{Compilation of the performance values for the different observing modes for the gamma-hadron separation cuts used in this analysis.  Detection is defined as 5$\sigma$ using the ring background method \citep{Berge2007}.
``Raw'' and ``corrected'' sensitivity values correspond to the sensitivity based on the raw exposure (the total length of time the array is taking data, excluding periods removed for bad weather) and live time, respectively.}
\label{tab:SensitivityNumbers}
\end{table*}

\begin{figure}[htb]
\centering
\includegraphics[width=\linewidth]{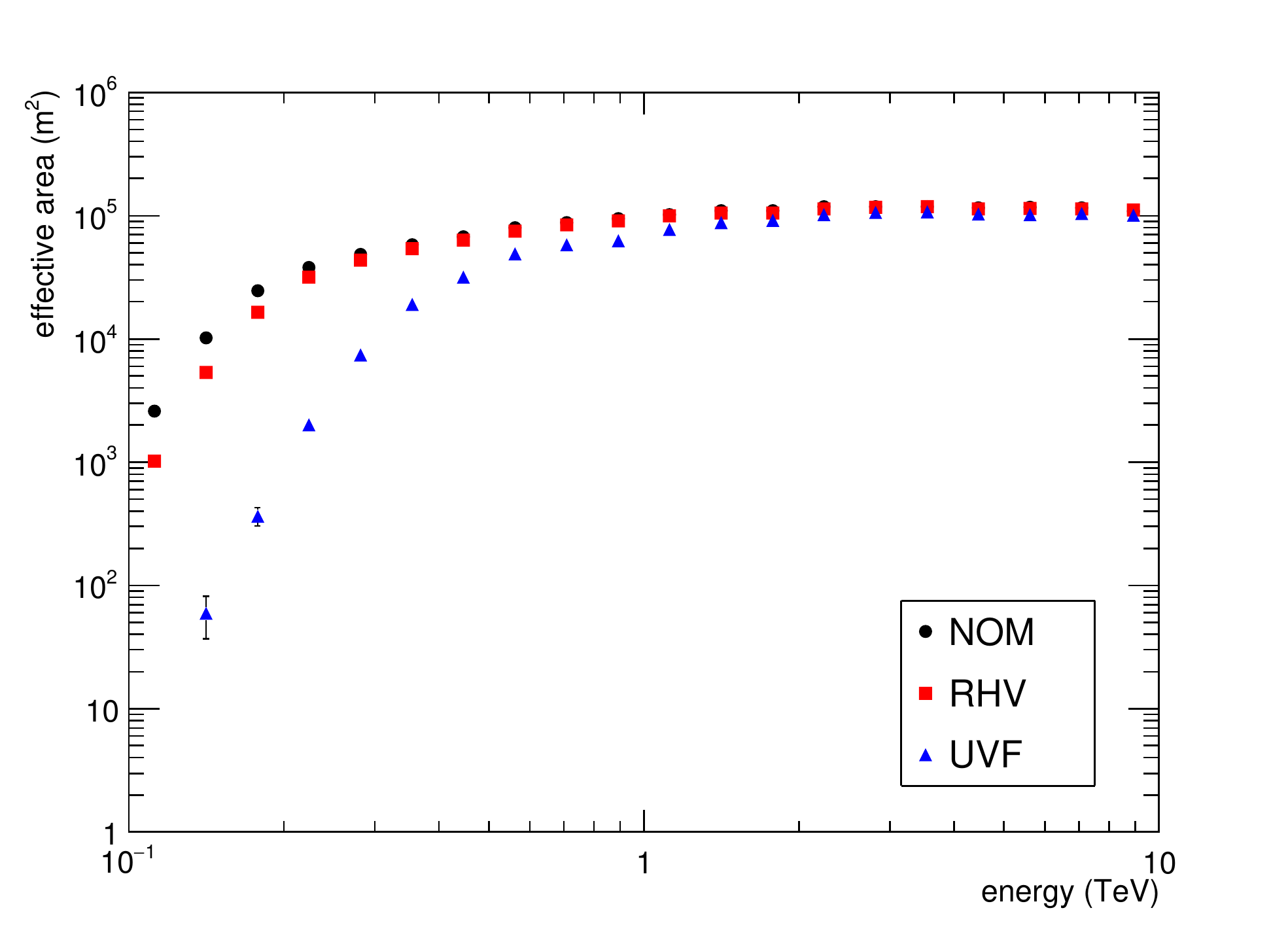}
\caption{Typical gamma-ray effective areas for the three observing modes at 20\deg\ zenith angle and typical observing conditions for that mode.
Above $300~\GeV$, RHV is very similar to NOM. 
Conversely, UVF is less than the NOM configuration at all energies; this is consistent with there being a reduced collection efficiency due to the reduction in the number of Cherenkov photons reaching the PMTs due to the filters, reducing the sensitivity to faint showers.}
\label{fig:EAs}
\end{figure}

With these analysis cuts, which were chosen to match the energy threshold of the NOM and RHV datasets, the NOM and RHV analyses have similar performance when considering the total exposure (time observing the source). 
This is in part due to the lower trigger rate of RHV observations ($180-200$~Hz \textit{cf.} 450~Hz for NOM) which reduces the deadtime of the array.
If a correction is made for this and the sensitivity is calculated as a function of the live time of the array, NOM observations are more sensitive than RHV.
The UVF configuration is significantly less sensitive than the NOM mode and has a higher analysis energy threshold and has typical trigger rates of about $50~\mathrm{Hz}$. 
Examples of the effective area as a function of energy used in this analysis are given in \cref{fig:EAs}.
The RHV effective area is comparable to the NOM effective area at all energies above about $300~\GeV$.  
Furthermore, for the cuts used in this work, the analysis-level energy threshold is similar for the two modes (about $160~\GeV$).  
The UVF energy threshold in this analysis is significantly higher (about $250~\GeV$) with a comparable effective area only reached above $1~\mathrm{TeV}$.

The angular resolution of each observing mode has also been investigated; the angular resolution of the RHV mode is similar to that of NOM (0.1$^\circ$ 68\% containment radius at 1~TeV). 
Conversely, UVF has systematically poorer angular resolution; this is a $15\%$ effect at $1~\mathrm{TeV}$ and a $35\%$ effect at $500~\mathrm{GeV}$.
This is consistent with there being less light arriving at the PMTs for a given shower which results in fewer PMTs contributing to the shower image, which increases the uncertainty in the reconstruction of the shower's arrival direction.

\section{Systematic Uncertainties Associated with New Observing Modes}
The dominant systematic uncertainties for IACTs are: limits in the knowledge of the time variability of the atmospheric parameters, the impact of degradation of the mirrors due to aging, and PMT-to-PMT variation in the quantum efficiency. 
None of these factors are impacted by the changes associated with these two new observing modes. 
Thus we do not expect a significant change in the systematic uncertainties. 
To estimate any additional impacts, we consider each mode in greater detail.

The PMT pulse shape and width vary as a function of high voltage, with the differences resulting from the voltage change used in this observing mode sufficiently small that it does not impact upon the telescopes operation and thus does not introduce any additional systematic error.
The additional PMT-to-PMT variation due to the reduction in gain is small, and is corrected for in the analysis of the data by using the nightly flasher runs and, when combined with other errors, they are not expected to increase the overall systematic uncertainty.
Comparisons of data to Monte Carlo simulations made for RHV show good agreement (comparable to those of the NOM data set). 
The standard estimate of the systematic uncertainty of the \veritas\ energy scale is $\sim 15-20 \%$ (see \cite{2015arXiv150807070P}), resulting in a systematic uncertainty of $20\%$ on the source flux and $0.2$ on the spectral index for a Crab Nebula-like spectrum.
Given the agreement between the RHV and NOM data/Monte Carlo comparisons, we maintain these systematic uncertainties for the RHV mode.

The impact on the systematic uncertainties is different for the UVF mode as the addition of the filters both adds additional material for reflection/absorption (the response of which can change over time like the rest of the telescope optics), and displaces the light cone entrance relative to the focal plane (increasing the optical point spread function).  
The effect on the telescope optics is small and is hence not expected to contribute to the overall systematic error of the measurement.

Over time, wear on the filter surfaces could degrade the overall transmittance of the filters (in the same way that the mirrors degrade).  
The mirror degradation is (conservatively) a $5\%$ per year effect in the overall reflectivity.  
However, unlike the mirror facets, the filters are not constantly exposed to the elements; they are shielded by the camera structure when in use and stored in a protective casing otherwise.  
Thus, aging of the filter surfaces is not expected to significantly impact the systematic uncertainties associated with UVF observations.

A source of additional uncertainty is due to the modeling of the optical properties of the filters.  
As was stated earlier, an attenuation factor has been determined using optical simulations of the \veritas\ focal plane.  
Any systematic difference between the calculated and true attenuation factor will have an effect on the absolute energy scale of the experiment, which will affect the overall systematic errors of the UVF observations.  
The analysis of UVF simulations has demonstrated that the energy reconstruction resolution of UVF data is systematically poorer than for NOM and RHV data at the energies relevant to this work; the magnitude of this effect is energy-dependent and ranges from a few percent (above 5~TeV) to about $20\%$ below 1~TeV.  
Based on this and the relative importance of this uncertainty in comparison to the other systematic uncertainties, we conservatively increase the estimate of the systematic uncertainties to $0.25$ and $25\%$ for the uncertainty on the spectral index and energy scale for a Crab Nebula-like spectrum, respectively. 

\section{Spectral Analysis of Crab Nebula Observations}
A useful check of new observing modes (and new instruments) is whether or not they reproduce a known result.
In VHE gamma-ray astrophysics, it is common to use the Crab Nebula for these purposes since it is a strong, steady point source at VHE energies. 

\begin{figure}[hbt]
\centering
\includegraphics[width=\linewidth]{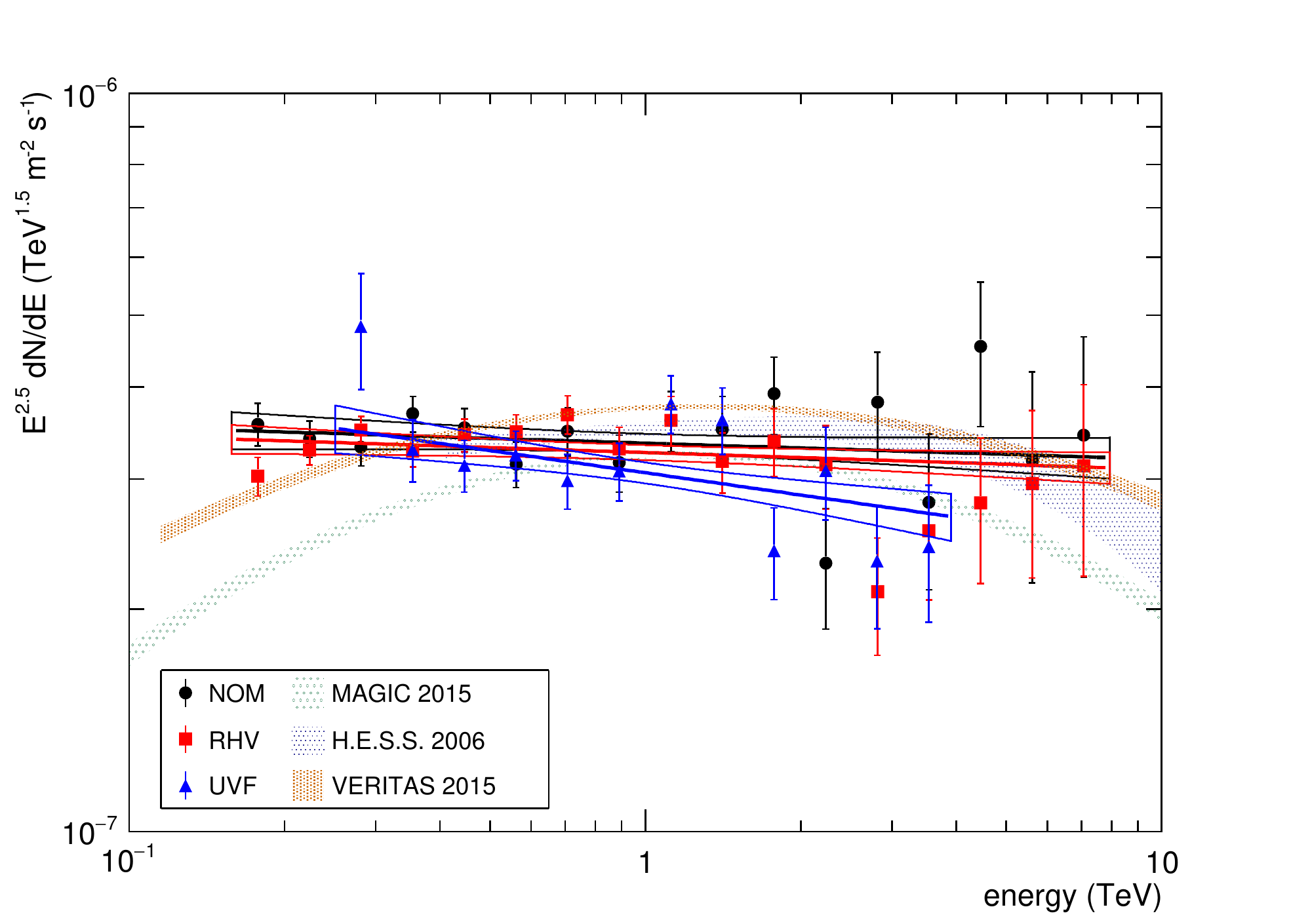}
\caption{Crab Nebula energy spectra for the three observing modes. 
Each data set has been fitted with a power law. 
The color bands represent the $1~\sigma$ contours on each fit (statistical errors only).
The values for all the fits are given in \cref{tab:CrabFitValues}.
Also included are historical measurements made by \hess\ \citep{2006A&A...457..899A}, \magic\ \citep{2015JHEAp...5...30A} and \veritas\ \citep{2015arXiv150806442K} shown as their 1~$\sigma$ statistical error bands.}
\label{fig:CrabSpectrum}
\end{figure}

The Crab Nebula's energy spectrum has been reconstructed for each observing mode using the same datasets as before. 
The data from each observing mode has been fitted with a power law, 
\begin{equation}
\frac{dN}{dE} = F_0 \left(\frac{E}{E_0}\right)^{-\Gamma};
\end{equation}
with the fitted values given in \cref{tab:CrabFitValues}, and plotted in \Cref{fig:CrabSpectrum} alongside the \veritas\ spectrum from \cite{2015arXiv150806442K}.
The NOM and RHV spectra agree well; the differences in the UVF spectra are within the uncertainties in the energy reconstruction coupled with fluctuations due to low statistics.
The spectra have also been compared with the published spectra from the \hess\ \citep{2006A&A...457..899A} and \magic\ \citep{2015JHEAp...5...30A} collaborations and agree within systematic errors.

\begin{table*}[hbt]
\centering
\begin{tabular}{ccccc} 
\hline\hline
\multirow{2}{*}{Dataset} & Energy Range & $F_0$  & \multirow{2}{*}{$\Gamma$} & \multirow{2}{*}{$\chi^2 / NDF$} \\ 
& (TeV) & ($10^{-7}\;\mathrm{m^{-2}~s^{-1}~TeV^{-1}}$) & & \\ \hline\hline 
NOM & 0.158 - 7.94 & $ 3.36 \pm 0.10_{stat}  \pm 0.67_{sys}$ & $ 2.52 \pm 0.03_{stat}  \pm 0.20_{sys} $  &  15.0 / 15  \\ 
RHV &  0.158 - 7.94 & $ 3.26 \pm 0.08_{stat}  \pm 0.65_{sys} $ & $ 2.52 \pm 0.02_{stat}  \pm 0.20_{sys} $ &  21.4 / 15  \\ 
UVF &  0.251 - 3.91 & $ 3.06 \pm 0.10_{stat}  \pm 0.77_{sys} $ & $ 2.60 \pm 0.05_{stat}  \pm 0.25_{sys} $ &  15.5 / 10   \\ \hline\hline 
\end{tabular}
\caption{Spectral fit results for the NOM, RHV, and UVF datasets presented in \cref{fig:CrabSpectrum}. Each dataset was fitted with a power law  at a normalization energy $E_0 = 1~\mathrm{TeV}$. }
\label{tab:CrabFitValues}
\end{table*}

\section{Application to Existing Observations and Future Projects}
Through these new modes, \veritas\ has been able to conduct both novel measurements and increase the exposure for existing observing programs.
The ability to monitor targets during moonlight has resulted in the detection of a flare from the blazar 1ES 1727+502 \citep{2015ApJ...808..110A} that otherwise would have gone unnoticed and significantly increased the exposure on a sequence of flares from the blazar 1ES~1959+650 \citep{2015ATel.8148....1M, 2016ATel.9010....1B}.
Furthermore, it is now easier to monitor periodic sources, such as the X-ray binary \textit{LSI~61\deg~303}, which has an orbital period close to the lunar cycle (26.5 days \cite{2002ApJ...575..427G}), which tends to restrict observations to roughly the same orbital phase. 
This object is known to undergo flaring \citep{2041-8205-817-1-L7} and shows orbital and super-orbital variability \citep{2013arXiv1308.0050A,2013ApJ...773L..35A}.
Also, the increased exposure allows for more flexible scheduling of multiwavelength observations.

An effort has also been made to measure the cosmic ray Moon shadow using the Earth-Moon ion spectrometer (EMIS) technique \citep{2001APh....14..287P} to measure the cosmic-ray positron fraction.
It is difficult to discriminate between positrons, electrons, and diffuse gamma rays due to the fact that the air showers caused by these particles is entirely electromagnetic and hence look identical in a standard analysis (whereas the majority of hadronic showers are easier to discriminate from electromagnetic ones).

A measurement can be conducted by observing the deficit in the cosmic ray flux (which is otherwise uniform) caused by the Moon. 
Since electrons and positrons are oppositely charged, their deficits are deflected in opposite directions due to the Earth's magnetic field.
The charge of the cosmic ray primary can then be measured since the amount of deflection is inversely proportional to the rigidity (momentum / charge).
When combined with standard analysis techniques to reconstruct the originating particle's properties it is possible to measure the relative deficit of different cosmic ray primaries and from this their relative abundances can be determined.
In particular this technique has the potential to measure the positron fraction ($e^+ / ( e^+ + e^-)$) and the antiproton ratio ($\bar{p} / p$)  above 500~GeV.

The EMIS technique requires measuring the cosmic ray flux close to (around 1-2\deg\ away from) a high elevation, partially illuminated Moon.
Using the \veritas\ telescopes, this can only be done by operating in both RHV and UVF modes simultaneously; for more information see \citep{Bird2015a}.

\section{Conclusions}
Two new observing modes for \veritas\ have been presented which increase the available observing time above the typical 1000 hours per year.
When operating with both RHV and UVF modes, the RHV mode provides a $13\%$ boost in yearly exposure above $160~\GeV$ and the UVF mode provides an additional $16\%$ boost in yearly exposure above $250~\GeV$ (a combined 30\% increase using the two modes).
Due to the additional effort involved in installing and removing the filters for UVF mode and an improved understanding of the RHV mode coupled with optimized target selection, it was deemed preferable to maximize the amount of RHV data taken rather than take data in both the RHV and UVF observing modes. 
Hence, only RHV data were taken during the 2014/15 and 2015/16 seasons. 
This has resulted in a 26\% increase in observing time over each of those seasons, which is only slightly less than the 30\% increase that was achieved in the 2013/14 observing season using both modes.
Note that in the event of (for example) a particularly spectacular astronomical event, UVF observations could still be conducted.

For the analysis cuts used in this work, the RHV mode gives comparable sensitivity to normal observations, and with a comparable analysis energy threshold (though it has a higher trigger energy threshold). 
In contrast, UVF has a reduced sensitivity, especially at low energies, and a raised energy threshold.
However, it is important to note that despite this, it is still useful observing time.

This additional observing time has been used to increase the live time of the experiment, this allows for deeper exposures, triggering or following up on astrophysical transient events, and facilitating the pursuit of new science goals.

\section*{Acknowledgements}
This research is supported by grants from the U.S. Department of Energy Office of Science, the U.S. National Science Foundation and the Smithsonian Institution, and by NSERC in Canada.  
We acknowledge the excellent work of the technical support staff at the Fred Lawrence Whipple Observatory and at the collaborating institutions in the construction and operation of the instrument. 
 R. Bird is funded by the DGPP which is funded under the Programme for Research in Third-Level Institutions and co-funded under the European Regional Development Fund (ERDF).  
 The \veritas\ Collaboration is grateful to Trevor Weekes for his seminal contributions and leadership in the field of VHE gamma-ray astrophysics, which made this study possible.

%
% References here
%
\section*{References}
\bibliography{brightMoonlight}

\end{document}